\documentclass[aps,eqsecnum,nofootinbib,superscriptaddress,showpacs]{revtex4}
\usepackage[dvips]{graphicx}  
\usepackage{amsmath,amsfonts}
\usepackage{bm}
\usepackage{color}  
\newcommand{\half}{ \frac{1}{2} }
\newcommand{\halfsqrt}{ \frac{1}{\sqrt{2}} }

\newcommand{\hx}{ \Hat{x} }

\newcommand{\ket}[1]{\left|~#1~\right\rangle}

\begin{document}
\title{Closed Superstrings in a Constant Magnetic Field \\
and \\
Regularization Criterion }
\author{Akira Kokado}
\email{kokado@kobe-kiu.ac.jp}
\affiliation{Kobe International University, Kobe 658-0032, Japan}
\author{Gaku Konisi}
\email{konisigaku@nifty.com}
\affiliation{Department of Physics, Kwansei Gakuin University,
Sanda 669-1337, Japan}
\author{Takesi Saito}
\email{tsaito@k7.dion.ne.jp}
\affiliation{Department of Physics, Kwansei Gakuin University,
Sanda 669-1337, Japan}
\date{\today}
\begin{abstract}
We propose a new type of interaction of closed superstrings with the electromagnetic field, other than 
the usual Kaluza-Klein type or a gauge field with internal gauge group origin. This model with a constant magnetic field 
is also shown to have an exact solution. We consider a regularization criterion. Some models will be excluded according 
to this criterion. The spectrum-generating algebra is also constructed in our interacting model.
\end{abstract}
\pacs{}
\maketitle
\section{Introduction}\label{sec:intro}
There is a long history of string models with electromagnetic interactions \cite{ref:Matsuda_S}-\cite{ref:Kokado_KS}. 
Recently a lot of interest has been drawn in exact solutions of closed strings placed in a uniform magnetic field 
\cite{ref:Russo_A}-\cite{ref:Chizaki_Y}. We can see the Landau-like energy level in these solutions. In these models 
the electromagnetic field has so far been introduced as a Kaluza-Klein type or a gauge field with internal gauge group origin. \\
\indent In the present paper we propose another type of interaction of closed superstrings with the electromagnetic field, 
which is introduced more directly in a gauge-invariant way (see Eq.(\ref{eq:intro_A})). At first sight this coupling seems 
to violate the superconformal invariance, as long as the superstring field $\hx^\mu $ is a superconformal scalar.  However, this is not 
the case when $\hx^\mu $ behaves in another way as we note in Eq.(\ref{eq:SCT3}). This model with a constant magnetic field is also shown 
to have an exact solution.\\
\indent The second aim of this paper is to consider a regularization criterion. Some models will be excluded 
according to this criterion. \\
\indent As the third aim we give the spectrum-generating algebra (SGA) in our interacting model. Physical states satisfying 
Virasoro conditions or equivalently the BRST charge condition are actually constructed from spectrum-generating operators. \\
\indent In Sec.\ref{sec:2} we propose a new type of interaction of the closed NSR superstring with the electromagnetic field. 
In Sec.\ref{sec:3} and Sec.\ref{sec:4} we calculate anomalies associated with the super Virasoro algebra. In order to calculate anomalies we use 
the operator product expansion method. However, we give also other regularization methods in Appendices A, B and C, 
in order to emphasize uniqueness of regularization.  In Sec.\ref{sec:5} the spectrum-generating algebra in our model is constructed. 
In Sec.\ref{sec:6} we consider an algebra isomorphic to the spectrum-generating algebra. From this we derive the number of space-time dimensions 
together with normal ordering constants of the 0-th Virasoro operator. In Sec.\ref{sec:7} and Sec.\ref{sec:8}, exact solutions of the heterotic string 
in a constant magnetic field are considered for both cases, one is ours and the other is of the KK type. We conclude that 
both exact solutions are unfortunately excluded according to the regularization criterion.  
In Sec.\ref{sec:9} the energy spectrum of NSR superstring in the constant magnetic field is summarized. 
In Sec.\ref{sec:10} we consider the regularization criterion. Finally Sec.\ref{sec:11} is devoted to concluding remarks.  \\
\indent  Appendices are composed of three regularization methods other than the operator product expansion method: \\
\indent Appendix A  Calculation of anomalies based on contraction \\
\indent Appendix B  Uniqueness of anomalies based on the damping factor method \\
\indent Appendix C  Regularization by means of the generalized Zeta function of Riemann
  
\section{A closed superstring in a constant magnetic field}\label{sec:2}
The action for free closed superstring is given by
\begin{align}
 &  S^0 = \int d^2sd\theta d\bar{\theta }~\hat{L}^0~, \quad \hat{L}^0 = 2\bar{D}\hx_\mu D\hx^\mu ~.
  \label{eq:Lag1}
\end{align}
where $D=i\partial _\theta +\theta \partial $, $\partial =\frac{\partial }{\partial s}$, $s=\tau -\sigma $,
$\bar{D}=i\partial _{\bar{\theta }}+\bar{\theta }\bar{\partial }$, $\bar{\partial }=\frac{\partial }{\partial \bar{s}}$, 
$\bar{s}=\tau + \sigma $, \\
and
\begin{align}
 & \hat{x}^\mu (s, \bar{s}, \theta , \bar{\theta }) = x^\mu (s, \bar{s}) 
 + i\frac{1}{\sqrt{2}}\theta \psi^\mu (s, \bar{s}) + i\frac{1}{\sqrt{2}}\bar{\theta }\bar{\psi}^\mu (s, \bar{s}) 
 + i\theta \bar{\theta }B^\mu (s, \bar{s})~.
  \label{eq:def_x}
\end{align}
If the action is integrated over $\theta , \bar{\theta }$, we have
\begin{align}
 & S^{0} = 2\int~d^{2}s~\Big[ \bar{\partial }x^{\mu }\partial x_{\mu } + i\half\big(\psi ^{\mu }\bar{\partial }\psi_{\mu } 
        + \bar{\psi }^{\mu }\partial \bar{\psi }_{\mu } \big) +B^\mu B_\mu \Big]~.
  \label{eq:def_S0}
\end{align}

Now, the electromagnetic field $A_\mu (\hat{x})$ is introduced in such a way as
\begin{align}
 & \hat{L} = 2[\bar{D}\hat{x}_\mu + 2\bar{\theta }qA_\mu (\hat{x})]D\hat{x}^\mu ~,
 \label{eq:intro_A}
\end{align}
where 2$q$ is regarded as a charge density along the closed string.  If the interaction term 
is integrated over $\theta ,\bar{\theta }$, we get
\begin{align}
 & S_{int} = 4q\int d^2s d\theta d\bar{\theta} \bar{\theta }A_\mu (\hat{x})D\hat{x} \nonumber \\
 & = 4q\int d^2s d\theta d\bar{\theta} [\bar{\theta }A_\mu (x) 
    + i\frac{1}{\sqrt{2}}\bar{\theta }\theta \psi ^\nu \partial_\nu A_\mu (x)]
   [-\frac{1}{\sqrt{2}}\psi ^\mu + \theta \partial x^\mu  
    - i\frac{1}{\sqrt{2}} \bar{\theta }\theta \partial\bar{\psi }^\mu + i\bar{\theta }B^\mu ] 
  \label{eq:intL1} \\
 & = 4q\int d^2s [A_\mu (x) \partial x^\mu  - i\frac{1}{2}\partial _\mu A_\nu (x)\psi ^\mu \psi ^\nu ] \nonumber 
\end{align}
The field equations say that $B^\mu =0$, so it is legitimate to simply set $B^\mu $ to zero, 
and  henceforth we forget it. Note that the superfield $\hat{x}$ cannot be divided into right- 
and left-moving sectors, because of introducing electromagnetic interaction. The coupling$A_\mu (\hat{x})D\hat{x}$, 
therefore, does not mean that $A_\mu (\hat{x})$ is coupled only with the right-moving mode. \\
\indent The interaction is clear to be invariant under the gauge transformation,
\begin{align}
 & \delta A_\mu (\hat{x}) = \partial _\mu \Lambda (\hat{x})~.
 \label{eq:gaugetrans_A}
\end{align}
We then choose the symmetric gauge
\begin{align}
 & A_{\mu }(\hx ) = - \half F_{\mu \nu } \hx^{\nu }~, \quad F_{\mu \nu } = \mbox{constant}~,
 \label{eq:sysgauge}
\end{align}
to obtain
\begin{align}
 & \Hat{L} = 2\big[\bar{D}\hx - \bar{\theta }qF \cdot \hx \big]\cdot D \hx ~.
 \label{eq:L2}
\end{align}
\indent If we use a new variable
\begin{align}
 & \Hat{X} = \exp(-qF\bar{s})\hx ~,
 \label{eq:newvariable}
\end{align}
we get
\begin{align}
 & \bar{D}\hat{X} = \big( i\partial _{\bar{\theta }}+\bar{\theta }\bar{\partial }\big) \big[\exp(-qF\bar{s})\hx\big] 
                 = \exp(-qF\bar{s})\big[-\bar{\theta }qF\hx + i\partial _{\bar{\theta }}\hx + \bar{\theta }\bar{\partial }\hx\big] 
                 = \exp(-qF\bar{s})\big[\bar{D}\hx - \bar{\theta }qF\hx \big]~, \nonumber \\
 & D\hat{X} = \big( i\partial _{\theta }+\theta \partial \big) \big[\exp(-qF\bar{s})\hx\big] 
                 = \exp(-qF\bar{s})D\hx~. \nonumber
\end{align}
From this observation we find that the Lagrangian (\ref{eq:L2}) reduces to
\begin{align}
 &  \Hat{L} = 2 \bar{D} \Hat{X}\cdot D \Hat{X}~.
  \label{eq:Lagrangian2}
\end{align}
This is a free-type Lagrangian with respect to $\Hat{X}$. The action for Eq.(\ref{eq:Lagrangian2}) is invariant under SCT, 
if $\hat{X}$ is a superconformal scalar. In this case the original variable $\hx$ behaves as
\begin{align}
 & \delta _{\mbox{SCT}}~\hx = \delta \bar{s}~qF\cdot \hx
 \label{eq:SCT3}
\end{align}
under SCT. At this point our approach is completely different from others [5][6], 
where $\hx$ is kept to be a superconformal scalar.  \\
\indent Let us concentrate on one of $2\times 2$ block of $F_{\mu \nu }$ with $B$ real
\begin{align}
  F_{\mu \nu }
  = \begin{pmatrix}
              0 &  B \\
             -B & 0
    \end{pmatrix}~, \quad \mu, \nu = 1, 2~.
 \label{eq:defF}
\end{align}
Introducing complex variables
\begin{align}
 &  \Hat{X}^{(\pm)} = \big( \Hat{X}^{1} \pm i\Hat{X}^{2}\big)/\sqrt{2}~.
 \label{eq:defX_pm}
\end{align}
Eqs.(\ref{eq:newvariable}) turns out to be
\begin{align}
 & \Hat{X}^{(\pm)} = \exp{(\pm iqB\bar{s})}\hx^{(\pm)}~.
 \label{eq:X_pm2}
\end{align}
Corresponding to Eq.(\ref{eq:def_x}), we expand $\Hat{X}^{(\pm)}(\tau ,\sigma ,\theta , \bar{\theta})$ 
in $\theta , \bar{\theta}$
\begin{align}
 & \Hat{X}^{(\pm)}(\tau ,\sigma ,\theta , \bar{\theta}) = X^{(\pm)}(\tau ,\sigma ) 
                 + i\frac{1}{\sqrt{2}}\theta\chi ^{(\pm )}(\tau ,\sigma )
                 + i\frac{1}{\sqrt{2}}\bar{\theta}\bar{\chi }^{(\pm )}(\tau ,\sigma )~.
 \label{eq:expandX_pm1}
\end{align}
From Eq.(\ref{eq:X_pm2}) we find
\begin{align}
 & X^{(\pm)}(\tau ,\sigma ) = \exp{\big[ \pm iqB(\tau + \sigma )\big]}x^{(\pm)}(\tau , \sigma )~, 
 \label{eq:Xpm3} \\
 & \chi ^{(\pm )}(\tau ,\sigma ) = \exp{\big[ \pm iqB(\tau + \sigma )\big]}\psi ^{(\pm)}(\tau , \sigma )~,
 \label{eq:Psi3} \\
 & \bar{\chi }^{(\pm )}(\tau ,\sigma ) = \exp{\big[ \pm iqB(\tau + \sigma )\big]}\bar{\psi }^{(\pm)}(\tau , \sigma )~, 
\end{align}
Considering the periodicity of $x^{(\pm)}(\tau , \sigma )$, $\psi ^{(\pm)}(\tau , \sigma )$ and
$\bar{\psi }^{(\pm)}(\tau , \sigma )$, 
we have the quasi-periodicity for $X^{(\pm)}(\tau ,\sigma )$, $\chi ^{(\pm )}(\tau ,\sigma )$ and 
$\bar{\chi }^{(\pm )}(\tau ,\sigma )$ as
\begin{align}
 & X^{(\pm)}(\tau ,\sigma + 2\pi ) = \exp{\big( \pm 2\pi iqB\big)}X^{(\pm)}(\tau , \sigma ) 
                                   = \exp{\big( \pm 2\pi i\omega \big)}X^{(\pm)}(\tau , \sigma )~, 
 \label{eq:Xperiodicity} \\
 & \chi ^{(\pm)}(\tau ,\sigma + 2\pi ) = \mp\exp{\big( \pm 2\pi i\omega \big)}\chi ^{(\pm)}(\tau , \sigma )~, 
 \label{eq:Psiperiodicity_chi} \\
 & \bar{\chi }^{(\pm)}(\tau ,\sigma + 2\pi ) = \mp\exp{\big( \pm 2\pi i\omega \big)}\bar{\chi }^{(\pm)}(\tau , \sigma )~, 
 \label{eq:Psiperiodicity_bar_chi}
\end{align}
where where |and + signs in front of exponential functions stand for NS sector and Ramond sector, respectively, 
and $qB=N+\omega $, $N \in Z$, $0\leq \omega < 1$. In the following we call $\omega $ simply the cyclotron frequency, 
and consider only the range $0<\omega <1$, since our system is equivalent to the free system when $\omega =0$. 
It is remarkable that when  $qB$  takes integral values $B=N\in Z$, , our system is equivalent to the completely free system.
There are no true anomalies in existing literatures, which should be written by $\omega $.\\
\indent The Lagarangian (\ref{eq:Lagrangian2}) can be written as
\begin{align}
 &  \Hat{L} = 2\big[ \bar{D}\Hat{X}^{(+)} D\Hat{X}^{(-)} + \bar{D}\Hat{X}^{(-)} D\Hat{X}^{(+)} \big ] 
             + 2 \sum_{\mu \neq 1,2}\bar{D}\Hat{x}^\mu  D\hx_\mu ~.
 \label{eq:redefL}
\end{align}
The first two terms are interacting terms, whereas the last term is free and well known.
Integrating $\Hat{L}$ over $\theta ,\bar{\theta }$, we have
\begin{align}
 & L=2\Big[ \bar{\partial }X^{(+)} \partial X^{(-)} + \bar{\partial }X^{(-)} \partial X^{(+)} 
    + \frac{i}{2}\big( \chi ^{(+)}\bar{\partial }\chi ^{(-)} + \chi ^{(-)}\bar{\partial }\chi ^{(+)}
    +\bar{\chi }^{(+)}\partial \bar{\chi }^{(-)} + \bar{\chi }^{(-)}\partial \bar{\chi }^{(+)} \big)\Big] \nonumber \\
 &   + 2\Big[\sum_{\mu \neq 1,2}\bar{\partial }x^{\mu }\partial x_\mu + \frac{i}{2}\big( \psi ^{\mu }\bar{\partial }\psi _{\mu } 
    + \bar{\psi }^{\mu }\partial \bar{\psi }_{\mu } \big) \Big]~.
 \label{eq:L3}
\end{align}
In spite of the quasi-periodicity of $X^{\pm}$, $\chi^{(\pm)}$ and $\bar{\chi}^{(\pm)}$, the periodic boundary condition, 
which is necessary in the variational principle, is guaranteed, because the aperiodic phase factors $\exp{(\pm 2\pi i\omega )}$ 
are always canceled out between the $(+)$ and $(-)$ components in the Lagrangian.  \\
\indent Equations of motion are all of free type:
\begin{align}
 &  \bar{\partial }\partial X^{(\pm)} = 0~,
 \label{eq:eq_of_motionX} \\
 &  \bar{\partial}\chi ^{(\pm)} = 0, \quad \partial\bar{\chi }^{(\pm)} = 0 ~.
 \label{eq:eq_of_motionPsi}
\end{align}
Their solutions with boundary conditions (\ref{eq:Xperiodicity}), (\ref{eq:Psiperiodicity_chi}) and 
(\ref{eq:Psiperiodicity_bar_chi}) are given by
\begin{align}
 & X^{(\pm)}(\tau ,\sigma ) = {X_{R}}^{(\pm)}(s) + {X_{L}}^{(\pm)}(\bar{s}) ~,
  \label{eq:expandX2} \\
 & {X_{R}}^{(\pm )}(s) = i \halfsqrt\sum _{n}\frac{1}{n\pm \omega } \exp{[-i(n\pm \omega )s]} {\alpha _n}^{(\pm)}~,
  \nonumber \\
 & {X_{L}}^{(\pm)}(\bar{s}) = i \halfsqrt\sum _{n}\frac{1}{n\mp \omega } \exp{[-i(n\mp \omega )\bar{s}]} {\tilde {\alpha }_n}^{(\pm)}~,
  \nonumber
\end{align}
and
\begin{align}
 & \chi ^{(\pm)}(s) = \sum _{r\in Z+\half}{b_r}^{(\pm)}\exp{[-i(r\pm \omega )s}~, \quad \mbox{for NS sector}
  \label{eq:expandX3} \\
 & \chi ^{(\pm)}(s) = \sum _{n\in Z}{d_n}^{(\pm)}\exp{[-i(n\pm \omega )s]}~, \quad \mbox{for Ramond sector}~.
  \nonumber \\
 & \bar{\chi }^{(\pm)}(\bar{s}) = \sum _{r\in Z+\half}{\tilde {b}_r}^{(\pm)}\exp{[-i(r\pm \omega )\bar{s}]}~, 
 \quad \mbox{for NS sector}
  \label{eq:expandX4} \\
 & \bar{\chi }^{(\pm)}(\bar{s}) = \sum _{n\in Z}{\tilde{d}_n}^{(\pm)}\exp{[-i(n\pm \omega )\bar{s}]}~, 
\quad \mbox{for Ramond sector}~.
  \nonumber
\end{align}
\indent  The conjugate momenta to $X^{(\pm)}(\tau , \sigma )$ are
\begin{align}
 &  P^{(\mp)}= \frac{\partial L}{\partial \big(\partial _\tau X^{(\pm)}\big )} = \partial X^{(\mp)} + \bar{\partial }X^{(\mp)} 
             = \dot {X}^{(\mp)}~.
 \label{eq:defP}
\end{align}
The quantization is accomplished by setting the commutation rules
\begin{align}
  & \big[\, X^{(\pm)}(\tau , \sigma )\,,~P^{(\mp)}(\tau , \sigma ')\, \big] = 2i\, \pi \delta _{\pm\omega }(\sigma -\sigma ')~,
\label{eq:def-XP-CCR}
\end{align}
and other combinations are zero. Here $\delta _{\pm\omega }(\sigma -\sigma ')$ is the delta function 
with the same quasi-periodicity as $X^{(\pm)}(\tau , \sigma )$  with respect to its argument. 
From these we have commutation relations:
\begin{align}
 & \big[\, {\alpha }_m^{(+)}\,,~{\alpha }_n^{(-)}\, \big] = (m+\omega )\delta _{m+n,0}~, \quad 0<\omega <1
\label{eq:def_alpha=ccR} \\
 & \big[\, {\tilde {\alpha }}_m^{(+)}\,,~{\tilde {\alpha }}_n^{(-)}\, \big] = (m-\omega )\delta _{m+n,0}~,
\label{eq:def-beta-CCR}
\end{align}
and other combinations are zero. Since $0<\omega <1$, $\alpha _m^{(-)}$, $\tilde {\alpha }_m^{(+)}$  are annihilation operators 
for $m>0$, and creation operators for $m\leq 0$, while $\alpha _m^{(+)}$, $\tilde {\alpha }_m^{(-)}$  are creation operators for $m<0$, 
and annihilation operators for $m\geq 0$. \\
\indent As for the fermionic parts, in the same way we get, for right-moving modes
\begin{align}
 & \big\{\, b_r^{(+)}\,,~b_s^{(-)}\, \big\} = \delta _{r+s,0}~, \quad \mbox{others }=0~,
\nonumber \\
 & \big\{\, d_m^{(+)}\,,~d_n^{(-)}\, \big\} = \delta _{m+n,0}~, \quad \mbox{others }=0~.
\label{eq:def-d-CCR}
\end{align}
${b_r}^{(\pm)}$ are annihilation operators for $r>0$, and creation operators for $r<0$. 
The same is true for left-moving modes.  For the Ramond sector, we need a special care on the 0-modes, so the detail will be discussed in Sec.\ref{sec:4}. \\
\indent The full Virasoro operators are almost the same as those of the free Virasoro operators, but mode operators for 
$\mu ,\nu =1,2$, or (+), (|), defined as Eq.(\ref{eq:defX_pm}), are specially there in,
\begin{align}
 &  L_n = \half \sum_{m=-\infty }^{+\infty }:\big( \alpha _{-m}^{(+)} \alpha _{n+m}^{(-)} + \alpha _{-m}^{(-)} \alpha _{n+m}^{(+)} \big) :
        +\half \sum_{m=-\infty }^{+\infty } \sum_{\mu ,\nu \neq 1,2} : \eta _{\mu \nu } \alpha _{-m}^{\mu } \alpha _{n+m}^{\nu } :
   \nonumber \\
 &       +\half \sum_{r=-\infty }^{+\infty } \Big( r+\frac{n}{2}-\omega \Big) : b_{-r}^{(+)}b_{n+r}^{(-)} :
        +\half \sum_{r=-\infty }^{+\infty } \Big( r+\frac{n}{2}+\omega \Big) : b_{-r}^{(-)}b_{n+r}^{(+)} :
   \nonumber \\
 &   +\half \sum_{r=-\infty }^{+\infty } \sum_{\mu ,\nu \neq 1,2} \Big( r+\frac{n}{2} \Big) :\eta _{\mu \nu } b_{-r}^{\mu } b_{n+r}^{\nu } :
   \label{eq:VirasoroL_n} \\
 &  G_r = \sum_{n=-\infty }^{+\infty } \big( \alpha _{-n}^{(+)}b_{r+n}^{(-)} + \alpha_{-n}^{(-)} b_{r+n}^{(+)} \big) 
        + \sum_{n=-\infty }^{+\infty } \sum_{\mu ,\nu \neq 1,2} \eta _{\mu \nu } \alpha _{-n}^{\mu }b_{r+n}^{\nu }~.
   \nonumber
\end{align}
The left-moving Virasoro operators $\tilde {L}_n, \Tilde {G}_r$ are also the same as above, 
but mode operators should be replaced by tilded ones with $(+)\leftrightarrow (-)$, i.e., $\alpha _n^{(\pm)}\rightarrow \tilde{\alpha }_n^{(\mp)}$ 
and $b_r^{(\pm)}\rightarrow \tilde{b}_r^{(\mp)}$.    
\section{Calculation of anomaly} \label{sec:3}
The Lagrangian (\ref{eq:L3}) happens to appear as if it is a free type. However, the dynamical variables $X^{(\pm)}(\tau ,\sigma )$ and 
$\chi ^{(\pm)}(\tau ,\sigma ), \bar{\chi }^{(\pm)}(\tau ,\sigma )$  are subject to the quasi-periodicity (\ref{eq:Xperiodicity})-(\ref{eq:Psiperiodicity_bar_chi}), and this 
causes the inclusion of the cyclotron frequency ƒÖ in the commutators (\ref{eq:def_alpha=ccR}) and (\ref{eq:def-beta-CCR}) for mode operators, 
which are different from the completely free case. Considering this fact, we should examine the validity of the super Virasoro 
algebras together with their anomalies. \\
\indent It is enough to consider only the right-moving part. Let us define current operators for interacting parts by
\begin{align}
 & J^{(\pm)}(z) = i\partial_{z}X_{R}^{(\pm)}(z) = \halfsqrt z^{\mp\omega } \sum_n z^{-n-1} \alpha _n^{(\pm)} 
  = \halfsqrt z^{\mp\omega }J_{0}^{(\pm)}(z)~, \quad z=\exp{(is)}~,
 \label{eq:defJ}
\end{align}
with
\begin{align}
 & J_{0}^{(\pm)}(z) = \sum_n z^{-n-1} \alpha _n^{(\pm)}~.
\label{eq:defJ0}
\end{align}  
In the following we use the notation $\partial $ for the derivative $\partial_{z}=\partial /\partial z $ by omitting the index $z$. \\
\indent The operator product expansions for them are given by
\begin{align}
 &  J_{0}^{(+)}(z)J_{0}^{(-)}(z') = \frac{1}{(z-z')^2} + \frac{\omega }{z'(z-z')}~,
 \label{eq:J+J-} \\
 &  J_{0}^{(-)}(z)J_{0}^{(+)}(z') = \frac{1}{(z-z')^2} - \frac{\omega }{z(z-z')}~,
 \label{eq:J-J+} 
\end{align} 
Here, we have used the following contractions:
\begin{align}
 & \langle~\alpha _m^{(+)}\alpha _n^{(-)}~\rangle = \delta _{m+n,0} \theta _{m\geq 0}(m+\omega )~,
  \nonumber \\
 & \langle~\alpha _m^{(-)}\alpha _n^{(+)}~\rangle = \delta _{m+n,0} \theta _{m> 0}(m-\omega )~,
 \label{eq:VEV_alpha_alpha0} \\
 & \quad 0 < \omega < 1~, \quad  
   \theta _\Gamma = \left\{
    \begin{array}{rl}
     1,& \quad \mbox{if $\Gamma$ is true} \\
     0,& \quad \mbox{if $\Gamma$ is false}
    \end{array}\right.
 \nonumber 
\end{align}
For the fermionic fields we confine ourselves to the NS sector,
\begin{align}
 &  \chi ^{(\pm)}(z) = z^{\mp\omega } \sum _r z^{-r-1/2} {b_r}^{(\pm)} = z^{\mp \omega }\chi _{0}^{(\pm)}(z)~,
 \label{eq:def_chi} \\
 &  \chi _{0}^{(\pm)}(z)\chi _{0}^{(\mp)}(z') = \frac{1}{z-z'}~,
 \label{eq:_chi_chi2}
\end{align}
with contractions $\left\langle b_r^{(+)}b_s^{(-)} \right\rangle =\left\langle b_r^{(-)}b_s^{(+)} \right\rangle = \delta _{t+s,0}\theta _{r>0}$ . 
The exponent |1/2 on $z$ in Eq.(\ref{eq:def_chi}) is only for convenience. \\
\indent Define the super current operator for interacting parts by
\begin{align}
 &  G(z) = \sqrt{2} \sum^2_{\mu =1} \chi _{\mu }(z) J^{\mu }(z) 
         = \chi _{0}^{(+)}(z) J_{0}^{(-)}(z) + \chi _{0}^{(-)}(z)J_{0}^{(+)}(z)~.
 \label{eq:defG}
\end{align}
Then we calculate the operator product $G(z)G(z')$ to yield the conformal operator $T(z)$, i.e.,
\begin{align}
 &  G(z)G(z') = \frac{2}{(z-z')^3} + \frac{\omega }{zz'(z-z')} + \frac{2T(z')}{z-z'}~,
 \label{eq:GG'}
\end{align}
where
\begin{align}
 &  T(z) = T^B(z) + T^F(z) = \half\sum_{\mu =1}^{2}\big[ :J_{0\mu }J_{0}^{\ \mu }: + :\partial \chi _\mu \chi^\mu :\big] ~.
 \label{eq:defT}
\end{align}
Here we have used the formula for the fermionic part
\begin{align}
 & :\partial \chi _\mu \chi^\mu : = :\partial \chi^{(+)} \chi^{(-)} + \partial \chi^{(-)} \chi^{(+)} : \nonumber \\
 & \quad \quad = :\partial\big( z^{-\omega } \chi_{0}^{(+)}\big) z^{+\omega }\chi_{0}^{(-)} + \partial \big( z^{+\omega }\chi_0^{(-)}\big) z^{-\omega }\chi_{0}^{(+)} :  \label{eq:formula_fermion} \\
 & \quad \quad = :- \frac{2\omega }{z}\chi_{0}^{(+)} \chi_{0}^{(-)} + \partial \chi_{0}^{(+)} \chi_{0}^{(-)} 
+ \partial \chi_{0}^{(-)} \chi_{0}^{(+)} : ~. \nonumber
\end{align}
In the same way we get
\begin{align}
 &  T^B(z)T^B(z') = \frac{1}{(z-z')^4} + \frac{\omega - \omega ^2}{zz'(z-z')^2} 
 + \frac{2T^B(z')}{(z-z')^2} + \frac{\partial 'T^B(z')}{z-z'}~,
 \label{eq:TBT'B}
\end{align}
for the bosonic part  $T^{B} = (1/2)\sum_{\mu =1}^{2}:J_{0\mu }J_{0}^{\mu }:$, and
\begin{align}
   T^F(z)T^F(z') = \frac{1/2}{(z-z')^4} + \frac{\omega ^2}{zz'(z-z')^2} + \frac{2T^F(z')}{(z-z')^2}
 + \frac{\partial 'T^F(z')}{z-z'}~.
 \label{eq:TFTF'}
\end{align}
for the fermionic part $T^{F}=(1/2):\partial \chi \cdot\chi :$.  Totally, it follows that
\begin{align}
   T(z)T(z') = \frac{3/2}{(z-z')^4} + \frac{\omega }{zz'(z-z')^2} + \frac{2T(z')}{(z-z')^2}
 + \frac{\partial 'T(z')}{z-z'}~.
 \label{eq:TT'2}
\end{align}
It is remarkable that the $\omega ^2$ anomaly in each of the bosonic term in Eq.(\ref{eq:TBT'B}) and the fermionic term in Eq.(\ref{eq:TFTF'}) is canceled out with each other in the total equation in (\ref{eq:TT'2}). The algebra is closed by
\begin{align}
   T(z)G(z') = \frac{3/2}{(z-z')^2}G(z') + \frac{1}{z-z'}\partial 'G(z') ~.
 \label{eq:T-G'}
\end{align}
\indent We have so far considered only the $(1, 2)$ plane, where the constant magnetic field is placed. 
The other ($d-2$)space-time components of fields are all free, and their Virasoro algebras are well known. Collecting all of them, we get
\begin{align}
 &  T(z)T(z') = \frac{3d/4}{(z-z')^4} + \frac{\omega }{zz'(z-z')^2} + \frac{2T(z')}{(z-z')^2}
 + \frac{\partial 'T(z')}{z-z'}~,
 \label{eq:T+T+'2} \\
 &  T(z)G(z') = \frac{3/2}{(z-z')^2}G(z') + \frac{1}{z-z'}\partial 'G(z') ~.
 \label{eq:T-G'2} \\
 &  G(z)G(z') = \frac{d}{(z-z')^3} + \frac{\omega }{zz'(z-z')} + \frac{2T(z')}{z-z'}~.
 \label{eq:GG'2} 
\end{align}
These are equivalent to the super Virasoro algebra
\begin{align}
 & \big[\, L_{m}\,,~L_{n}\, \big] = (m-n)L_{m+n} + \delta _{m+n,0}A_m~,
\label{eq:L+L+CCR} \\
 & \big[\, L_{m}\,,~G_r\, \big] = \big(\frac{m}{2}-r\big)G_{m+r}~,
\label{eq:L-GrCCR} \\
 & \big\{\, G_r\,,~G_s\, \big\} = 2L_{r+s} + \delta _{r+s,0}B_r~,
\label{eq:GrGrCCR} 
\end{align}
where the anomaly terms are given by
\begin{align}
 &  A_m = \frac{d}{8}m(m^2-1) + m\omega~,
 \label{eq:Am+} \\   
 &  B_r = \frac{d}{2}\big(r^2-\frac{1}{4}\big) + \omega ~,
 \label{eq:Br-} \\
 &  qB = N+\omega , \quad N\in Z, \quad 0<\omega <1~.
 \label{eq:defB_q}
\end{align}
Anomalies for the left-moving part are the same as above, $i.e., \Tilde {A}_m=A_m, \Tilde {B}_r=B_r$. 
%
\section{Anomaly in the Ramond sector} \label{sec:4}
As for the Ramond sector, we should be careful for the 0-mode. The mode expansions of fermionic fields are given by
\begin{align}
   \chi _R^{(\pm)}(z) = z^{\mp \omega } \sum_n z^{-n}d_n^{(\pm)} = z^{\mp \omega } \chi _{R0}^{(\pm)}(z)~.
 \label{eq:chi_pm}
\end{align}
where $n$ runs over the integral-numbers. The mode operators obey the commutation relation, 
\begin{align}
  \big\{\, d_m^{(+)}\,,~d_n^{(-)}\, \big\} = \delta _{m+n,0}~.
\label{eq:d-CCR2}
\end{align}
Usually the 0-mode $d_0^{\mu }$ is regarded as the Dirac  $\gamma $-matrix. However, in the presence of the magnetic field, 
it is not the case. The reason is as follows: Note that the super Virasoro operator $F_0$ contains factors, 
$\alpha _0^{(+)}d_0^{(-)}+\alpha _0^{(-)}d_0^{(+)}$. Since $\alpha _0^{(-)}$ is the creation operator, 
the second term contradicts with the Virasoro condition $F_0\ket{\mbox{ground state}}=0$, if $d_0^{(\pm)}$ is regarded as the Dirac  $\gamma $ matrix. 
In the sector of the presence of magnetic fields, therefore, $d_0^{(+)}$ should be regarded as the annihilation operator, 
whereas other components $d_0^{\mu }$ without magnetic fields behave as $\gamma $ matrices. \\
\indent From this reason $d_m^{(+)}$ is regarded as annihilation operator for $m\geq 0$, 
and creation operator for $m<0$, while $d_m^{(-)}$ is annihilation operator for $m>0$, and creation operator for $m\leq 0$. 
The contractions are, therefore, defined as
\begin{align}
 & \left\langle~d_m^{(+)}d_n^{(-)} ~\right\rangle =
  \left\{\begin{array}{rl}
        \delta _{m+n,0},& \quad (m\geq 0) \\
        0,& \quad (m<0)
         \end{array}\right.  
 \quad
 & \left\langle~d_m^{(-)}d_n^{(+)} ~\right\rangle =
  \left\{\begin{array}{rl}
        \delta _{m+n,0},& \quad (m>0) \\
        0,& \quad (m\leq 0)
         \end{array}\right.
 \label{eq:d_d} 
\end{align}
\\ 
The operator product expansions for fermionic fields are, then, given by
\begin{align}
 & {\chi }_{R0}^{(+)}(z){\chi }_{R0}^{(-)}(z') = \frac{z}{z-z'}~,
 \label{eq:chi+chi-2} \\
 & \chi _{R0}^{(-)}(z)\chi _{R0}^{(+)}(z') = \frac{z'}{z-z'}~.
 \label{eq:chi-chi+2}
\end{align}
For the super operator, $F(z)=\chi _{R0}^{(+)}(z)J_0^{(-)}(z)+\chi _{R0}^{(-)}(z)J_0^{(+)}(z)$, 
we have 
\begin{align}
  \mbox{anomaly terms of } F(z)F(z')=\frac{z+z'}{(z-z')^3}~.
 \nonumber
\end{align}
From the formula
\begin{align}
  \big\{\, F_m\,,~F_n\, \big\} = \oint dz dz'~z^{m}{z'}^{n} F(z)F(z')~,
\label{eq:defFmFn}
\end{align}
it follows that
\begin{align}
 & \big\{\, F_m\,,~F_n\, \big\} = 2L_{m+n} + \delta _{m+n,0}B_m(\mbox{Ramond})~,
\label{eq:FmFn2} \\
 & B_m(\mbox{Ramond}) = \frac{d}{2}m^2~.
\end{align}
For the fermionic part $T^F=(1/2):\partial \chi _R\cdot \chi _R:$ with
\begin{align}
 & 2 T^F = :\partial {\chi _R}^{(+)}{\chi _R}^{(-)} + \partial {\chi _R}^{(-)}{\chi _R}^{(+)} : 
 \nonumber \\
 & = :-\frac{2\omega }{z} {\chi _{R0}}^{(+)}{\chi _{R0}}^{(-)} + \partial {\chi _{R0}}^{(+)}{\chi _{R0}}^{(-)} 
  + \partial {\chi _{R0}}^{(-)}{\chi _{R0}}^{(+)} :~.
 \nonumber
\end{align}
we have
\begin{align}
  \mbox{Anomaly parts of }T^F(z)T^F(z') = \frac{1}{(z-z')^4}\frac{z^2+z'^2}{4} + \frac{\omega ^2 - \omega }{(z-z')^2}~.
 \label{eq:anomalyTFTF'}
\end{align}
\indent For the bosonic part $T^B=:J_0^{(+)}J_0^{(-)}:$, we already had the product $T^B(z)T^B(z')$ before as
\begin{align}
  \mbox{Anomaly parts of } T^B(z)T^B(z') = \frac{zz'}{(z-z')^4} + \frac{\omega -\omega ^2}{(z-z')^2}~.
 \label{eq:anomalyTBTB}
\end{align}  
Here the equation has been multiplied by the factor $zz'$, in order to make it of the same power as the fermionic one. 
Then the total sum of the anomaly is given by
\begin{align}
 & \mbox{Anomaly of } \big[T^B(z) + T^F(z)\big]\big[T^B(z') + T^F(z')\big] 
 \label{eq:anomalyT+T} \\ 
 &   = \frac{1}{(z-z')^4}\big( zz' + \frac{z^2 + z'^2}{4} \big)~.
 \nonumber
\end{align}  
where $\omega, \omega^2$ terms are cancelled out from Eqs.(\ref{eq:anomalyTFTF'}) and (\ref{eq:anomalyTBTB}). 
This gives the anomaly term without the cyclotron frequency
\begin{align}
   A_m(\mbox{Ramond}) = \frac{d}{8}m^3~,
 \label{eq:Am2} \\
   B_m(\mbox{Ramond}) = \frac{d}{2}m^2~.
 \label{eq:Bm2}
\end{align}
The same is true for the left-moving part.
%
\section{Spectrum-generating algebra} \label{sec:5}
The SGA for interacting dimensions $\mu=1,2$, or, $(+), (-)$ is characterized by the cyclotron frequency $\omega $. 
We summarize it for the right-moving NS sector:
\begin{align}
 & \big[\, A_m^{(+)}\,,~A_n^{(-)}\, \big] = (m+\omega )\delta _{m+n,0}~, \quad
  \big\{\, B_r^{(+)}\,,~B_s^{(-)}\, \big\} = \delta _{r+s,0}~,  \quad
  \big[\, A_m^i\,,~B_r^j\, \big] = 0~,
 \nonumber \\
 & \big[\, A_m^{(\pm)}\,,~A_n^+\, \big] = (m\pm \omega )A_{m+n}^{(\pm)}~, \quad
  \big[\, B_r^{(\pm)}\,,~A_n^+\, \big] = \big( \frac{n}{2} + r \pm \omega \big)B_{r+n}^{(\pm)}~, 
 \label{eq:SGA1} \\
 & \big[\, A_m^{(\pm)}\,,~B_r^+\, \big] = (m\pm \omega )B_{m+r}^{(\pm)}~, \quad
  \big\{\, B_r^{(\pm)}\,,~B_s^+\, \big\} = A_{r+s}^{(\pm)}~, 
 \nonumber
\end{align}
\begin{align}
 & \big[\, A_m^{+}\,,~A_n^{+}\, \big] = (m-n )A_{m+n}^{+} + m^3\delta _{m+n,0}~,
 \nonumber \\
 & \big[\, A_m^+\,,~B_r^+\, \big] = \big(\frac{m}{2} - r \big)B_{r+s}^+~,
 \label{eq:SGA2} \\
 & \big\{\, B_r^+\,,~B_s^+\, \big\} = 2A_{r+s}^{+} + 4r^2\delta _{m+n,0}~.
 \nonumber
\end{align}
The sub-algebra (\ref{eq:SGA2}) is completely the same as that in Ref.\cite{ref:Brower_F}, so we omit to define $A_n^+$ and $B_r^+$ explicitly.  
They are composed of free operators with light-cone components.
Any operator in Eqs.(\ref{eq:SGA1}) and (\ref{eq:SGA2}) is commutable with the super Virasoro operator $G_r$. \\
\indent The new operators $A_n^{(\pm)}$ and $B_r^{(\pm)}$ in Eq.(\ref{eq:SGA1}) are defined as 
\begin{align}
 & A_n^{(\pm )} = \frac{1}{2\pi} \int_{-\pi }^{\pi } d\tau  A_n^{(\pm )}(\tau )V^{n\pm \omega }(\tau )~,
 \label{eq:defAmpm} \\
 & \quad A_n^{(\pm)}(\tau ) =  P^{(\pm)} - (n\pm \omega )\chi ^{(\pm)}\psi _{-}~, \nonumber \\
 & B_r^{(\pm)} = \frac{1}{2\pi} \int_{-\pi }^{\pi } d\tau  B_r^{(\pm)}(\tau )V^{r\pm \omega }(\tau )~,
 \label{eq:defBrpm} \\
 & \quad B_r^{(\pm)}(\tau ) = \chi ^{(\pm)}\big( 1-\frac{i}{2} \psi_-\partial _{\tau }\psi _- P_{-}^{-2} \big) P_{-}^{1/2} 
  - \psi _- P^{(\pm)}P_-^{-1/2}~, \nonumber
\end{align}
where
\begin{align}
 & V(\tau ) = :\exp{[iX_-(\tau )}]:~.  
\end{align}
Here, $X_{-}$, $P_{-}$ and $P^{(\pm)}$, as well as $\chi ^{(\pm)}, \psi _{-}$, are all right-moving operators defined by
\begin{align}
 & X_-(\tau ) = \sqrt{2}X_R^{-}(\tau )=x_- + \tau p_- + i\sum_{n \neq 0} n^{-1} \alpha _n^{-} e^{-in\tau }~, 
  \nonumber  \\
 & P_{\pm}(\tau )=\sqrt{2}\partial _\tau X_R^{\pm}(\tau ) = \sum_ne^{-in\tau }\alpha _n^{\pm}~, 
 \label{eq:mode_expand} \\
 & P^{(\pm)}(\tau )=\sqrt{2}\partial _\tau X_R^{(\pm)}(\tau ) = \sum_n e^{-i(n\pm\omega )\tau }\alpha _n^{(\pm)}~.
 \nonumber
\end{align}
The nude indices $\pm$ denote the light-cone components defined as $X_{\pm}=\kappa ^{\pm}(X^0 \pm X^{d-1})/\sqrt{2}$ 
with a real parameter $\kappa $. The dressed indices $(\pm)$ of $X_{\pm}=(X^1 \pm i X^2)/\sqrt{2}$ should be distinguished 
from the light-cone indices $\pm$. The new definitions for $A_m^{(\pm)}$ and $B_r^{(\pm)}$ reduce to the original ones proposed 
by Brower and Friedmann\cite{ref:Brower_F}, if the cyclotron frequency $\omega $ is set to be zero. \\
\indent The sub-algebra (\ref{eq:SGA2}) is completely the same as that in Ref.\cite{ref:Brower_F}, so we omit to define $A_n^+$ and $B_r^+$ explicitly. 
They are composed of free operators with light-cone components. The proof of our SGA (\ref{eq:SGA1}) is given by the same method as in Ref.\cite{ref:Kokado_KS2}.
%
\section{Isomorphisms} \label{sec:6}
The algebra (\ref{eq:SGA2}) is similar to the super Virasoro algebra for transverse operators
\begin{align}
 & \big[\, L_m^{T}\,,~L_n^{T}\, \big] = (m-n)L_{m+n}^T + A^{T}(m)\delta _{m+n,0}~,
 \nonumber \\
&  \big[\, L_m^{T}\,,~G_r^{T}\, \big] = \big( \frac{m}{2}-r)G_{m+r}^{T}~,
\label{eq:LTGrCCR} \\
&  \big\{\, G_r^{T}\,,~G_s^{T}\, \big\} = 2L_{r+s}^{T} + B^{T}(r)\delta _{r+s,0}~,
 \nonumber 
\end{align}  
where
\begin{align}
 &  A^{T}(m) = \frac{d-2}{8}m(m^2-1) + 2ma + m\omega ~,
 \label{eq:ATm} \\
 &  B^{T}(r) = \frac{d - 2}{2}\big (r^2 - \frac{1}{4} \big ) + 2a + \omega ~,
 \label{eq:BTm} \\
 &  qB = N+\omega , \quad N\in Z, \quad 0<\omega <1~.
 \label{eq:Beq2}
\end{align}
Here the superscript $T$ means that the operators are constructed from $L_m, G_r$, 
leaving oscillators with spacial components $\mu =1,2,\cdots, d-2$. The constant $a$ 
is included in $L_{m}^{T}$ as $-a\delta _{m,0}$. \\
\indent  The isomorphisms 
\begin{align}
  A_{m}^{+} \sim L_{m}^{T}, \quad B_{r}^{+} \sim G_r^{T}~,
 \label{eq:AmLTBrGT}
\end{align}
are completed, if there hold equations
\begin{align}
 &  A^{T}(m) = \frac{d-2}{8}(m^3-m) + 2ma + m\omega = m^3~,
 \label{eq:ATm2} \\
 &  B^{T}(r) = \frac{d - 2}{2}\big (r^2 - \frac{1}{4} \big ) + 2a + \omega =4r^2~,
 \nonumber
\end{align}
These two equations are consistent to give the solution,
\begin{align}
 &  d = 10~,
 \label{eq:d-} \\
 &  a = \half (1-\omega )~.
 \nonumber
\end{align}
As for the Ramond sector, we have $d^{R} = 10$ and $a^{R}=0$. \\
\indent  The isomorphisms (\ref{eq:AmLTBrGT}) are also extended to other components interacting 
with the magnetic field. The algebra (\ref{eq:SGA1}) is similar to
\begin{align}
 & \big[\, \alpha _m^{(+)}\,,~\alpha _n^{(-)}\, \big] = (m+\omega )\delta _{m+n,0}~, \quad
  \big\{\, b_r^{(+)}\,,~b_s^{(-)}\, \big\} = \delta _{r+s,0}~,  \quad
  \big[\, \alpha _m^i\,,~b_r^j\, \big] = 0~,
 \nonumber \\
&  \big[\, \alpha _m^{(\pm)}\,,~L_n^T\, \big] = (m\pm \omega )\alpha _{m+n}^{(\pm)}~, \quad
  \big[\, b_r^{(\pm)}\,,~L_n^T\, \big] = \big( \frac{n}{2} 
+ r \pm \omega \big)b_{r+n}^{(\pm)}~, 
 \label{eq:SGA1a} \\
& \big[\,\alpha _m^{(\pm)}\,,~G_r^T\, \big] = (m\pm \omega )b_{m+r}^{(\pm)}~, \quad
  \big\{\, b_r^{(\pm)}\,,~G_s^T\, \big\} = \alpha _{r+s}^{(\pm)}~, 
 \nonumber 
\end{align}
The isomorphisms are now completed by
\begin{align}
   A_{m}^{(\pm)} \sim \alpha _m^{(\pm)}, \quad B_r^{(\pm)} \sim b_r^{(\pm)}~.
 \label{eq:AmbetaBrbr}
\end{align}
The same conclusion is obtained for the left-moving part. \\
\indent Any physical state should satisfy the BRST condition  $Q_{\mbox{BRST}}\ket{\mbox{phys.}}=0$, 
or equivalently the super Virasoro conditions, 
\begin{align}
 & G_{r>0}\ket{\mbox{phys.}}=0~, \quad \big( L_{n\geq 0}-\delta _{n,0}a \big)\ket{\mbox{phys.}}=0~,
  \label{eq:physcond_Gr_Ln} \\
 & \Tilde {G}_{r>0}\ket{\mbox{phys.}}=0~, \quad \big(\tilde {L}_{n\geq 0}-\delta _{n,0}a\big)\ket{\mbox{phys.}}=0~,
  \label{eq:physcond_tilGr_tilLn}
\end{align}
for the NS sector with $a=(1-\omega )/2$, and 
\begin{align}
 & F_{n\geq 0}\ket{\mbox{phys.}}=0~, \quad L_{n\geq 0}\ket{\mbox{phys.}}=0~,
  \label{eq:physcond_Fr_Ln} \\
 & \Tilde {F}_{n\geq 0}\ket{\mbox{phys.}}=0~, \quad \tilde {L}_{n\geq 0}\ket{\mbox{phys.}}=0~,
  \label{eq:physcond_tilFr_tilLn}
\end{align}
for the Ramond sector . It is well known that such physical states can be constructed by using spectrum-generating operators.
%
%
%
%
%
%
%
%
%
%
%
\section{Related solvable models} \label{sec:7}
We have shown that the closed superstring placed in a constant magnetic field can be solved exactly. If the fermionic field $\psi ^\mu (\tau ,\sigma)$ 
 is neglected from our model, we have a closed bosonic string in the constant magnetic field. This provides also another exactly solvable model. 
In this case the Virasoro constraint constant is given by $a=1-(\omega -\omega ^2)/2$, and the space-time dimension is $d=26$. \\
\indent As the third possibility of exactly solvable models, we can consider the heterotic string in the constant magnetic field. 
This heterotic model is obtained from our model by replacing the left-moving fermion with the 32 Lorentz singlet Majorana-Weyl fermion 
$\lambda ^A(\tau ,\sigma ), A=1, \cdots, 32$. The Lagrangian is given by
\begin{align}
 & \hat{L} = 2[\bar{\partial }\hat{x}_\mu + 2qA_\mu (\hat{x})]D\hat{x}^\mu + 2i\theta \sum_{A=1}^{32} \lambda _{+}^A \partial _{-}\lambda _{+}^A~,
 \label{eq:intro_HL}
\end{align}
where $F^{\mu \nu }=\partial ^\mu A^\nu - \partial ^\mu A^\nu =$const. \\
\indent Unfortunately, however, this heterotic model contains inconsistency, and thereby it fails to be valid. The reason is as follows: 
The heterotic string in a constant magnetic field is characterized by the Virasoro constraint equations, two of which are given by \\
\begin{align}
 & \big( L_{0}- a \big)\ket{\mbox{phys.}}=\big(\tilde {L}_{0}-\tilde{a}\big)\ket{\mbox{phys.}}=0~,
  \label{eq:physcond_L0} 
\end{align}
where
\begin{align}
 & a=(1-\omega)/2, \quad \Tilde{a}=1-(\omega -\omega ^2)/2~,
  \label{eq:a_tild_a} \\
 & L_{0}=\frac{\bar{p}^2}{2}+N, \quad \bar{p}^2\equiv \sum_{\mu ,\nu \neq 1,2}\eta _{\mu \nu }p^\mu p^\nu ~,
  \label{eq:cond_L0} \\
 & \Tilde{L}_{0}=\frac{\bar{p}^2}{2}+\tilde{N},
\end{align}
with
\begin{align}
 &  N = \sum_{n=1 }^{+\infty }\big( \alpha _{-n}^{(+)} \alpha _{n}^{(-)} + \alpha _{-n}^{(-)} \alpha _{n}^{(+)} \big) 
       + \alpha _{0}^{(-)} \alpha _{0}^{(+)}
        + \sum_{n=1}^{+\infty } \sum_{\mu ,\nu \neq 1,2} \eta _{\mu \nu } \alpha _{-n}^{\mu } \alpha _{n}^{\nu }
   \nonumber \\
 &       +\sum_{r=1/2 }^{+\infty } \big( r-\omega \big) b_{-r}^{(+)}b_{r}^{(-)} 
        +\sum_{r=1/2 }^{+\infty } \big( r+\omega \big) b_{-r}^{(-)}b_{r}^{(+)} 
        + \sum_{r=1/2}^{+\infty } \sum_{\mu ,\nu \neq 1,2} \eta _{\mu \nu } r b_{-r}^{\mu } b_{r}^{\nu } 
   \label{eq:def_N} \\
 & \Tilde{N}= \sum_{n=1 }^{+\infty }\big( \Tilde{\alpha }_{-n}^{(+)} \Tilde{\alpha }_{n}^{(-)} 
        + \Tilde{\alpha }_{-n}^{(-)} \Tilde{\alpha }_{n}^{(+)} \big) 
        + \Tilde{\alpha }_{0}^{(+)} \Tilde{\alpha }_{0}^{(-)}
        + \sum_{n=1}^{+\infty } \sum_{\mu ,\nu \neq 1,2} \eta _{\mu \nu } \Tilde{\alpha }_{-n}^{\mu } \Tilde{\alpha }_{n}^{\nu }
        + \sum_{A=1}^{32} \sum_{r=1/2}^{+\infty } r\lambda _{-r}^A\lambda _{r}^A~.
   \label{eq:def_tild_N}
\end{align}
Here we concern the NS superstring for the right-moving sector. The $(\omega -\omega ^2)/2$ anomaly in Eqs.(\ref{eq:a_tild_a}) 
comes from the anomaly term in Eq.(\ref{eq:TBT'B}). \\
\indent Let us now consider the difference
\begin{align}
 & \big(L_{0}-\tilde {L}_{0}-a+\tilde{a}\big)\ket{\mbox{phys.}}=\big(N-\tilde {N}+\half +\half\omega ^2\big)\ket{\mbox{phys.}}=0~,
  \label{eq:physcond_L0_tild_L0}
\end{align}
The last equation fails to be valid, because the $\omega ^2$ term cannot be canceled by any eigenvalue 
of the number operator difference, $N-\tilde{N}$, which gives at most the first order of $\omega $. 
Therefore, there is no possibility of the solvable heterotic model with Lagrangian (\ref{eq:intro_HL}). \\
\indent From this observation, we conclude that the exact solution in NSR superstring exists in each of combined sectors, 
(NS-NS), (NS-R) and (R-R), because they have no $\omega ^2$ anomaly. 
%
\section{The Kaluza-Klein type interaction of the heterotic strings} \label{sec:8}
Omitting the left-moving internal fermionic fields, $\lambda ^A(\tau ,\sigma ), A=1, \cdots, 32$, 
the Lagrangian for heterotic string with the KK (0,1)-type interaction can be written as \cite{ref:Russo_A}
\begin{align}
 & \Hat{L} = 2\big[\bar{\partial }\Hat{x}_\mu + 2\bar{\partial }\Hat{u}A_\mu (\Hat{x})\big] D\Hat{x}^\mu ~,
 \label{eq:intro_KKL}
\end{align}
Here we have defined variables $\hat{u}=(\hat{y}-\hat{t})/\sqrt{2}$, $\hat{v}=(\hat{y}+\hat{t})/\sqrt{2}$, with $t=x^0$ 
and $y\equiv x^4$, so that
\begin{align}
 & \bar{\partial }\hx_\mu D\Hat{x}^\mu = -\bar{\partial }\Hat{t}D\Hat{t} + \bar{\partial }\Hat{y}D\Hat{y} + \cdots 
   = \bar{\partial }\Hat{u}D\Hat{v} + \bar{\partial }\Hat{v}D\Hat{u} + \cdots~.
  \label{eq:part_KKL}
\end{align}
Here $y\equiv x^4$ is the fifth component of space-time, and is considered as an internal coordinate. \\
\indent The Lagrangian is then rewritten as
\begin{align}
 & \Hat{L} = 2 \big[\bar{\partial }\Hat{u} D\Hat{v} + \bar{\partial }\Hat{v}D\Hat{u} + \cdots +2\bar{\partial }\Hat{u}A_\mu (\hx)D\hx^\mu \big]~.
 \label{eq:L_2}
\end{align}
This is analogous to the previous Lagrangian (\ref{eq:intro_HL}), but here $A_{\mu }(x)$ is the KK electromagnetic field, 
and is assumed to be independent of $y$. \\
\indent In a gauge, $A_\mu (\hx)=-F_{\mu \nu }\hx^\nu /2$, ($F_{\mu \nu }=$const.), Eq.(\ref{eq:L_2}) becomes
\begin{align}
 & \Hat{L} = 2\big[\bar{\partial }\hx_\mu - \bar{\partial }\Hat{u}F_{\mu,\nu }\hx^\nu \big] D\hx^\mu 
    \nonumber \\
 &         = 2 \big[\bar{\partial }\Hat{u} D\Hat{v} 
              + \bar{\partial }\Hat{v}D\Hat{u} + \cdots -\bar{\partial }\Hat{u}F_{\mu \nu } \hx^\nu D\hx^\mu \big] ~.
 \label{eq:intro_KKL2}
\end{align}
From this Lagrangian we have field equations
\begin{align}
 &  \bar{\partial }D \hx_\mu  - \half \bar{\partial }\Hat{u}F_{\mu \nu }D\hx^\nu 
       - \half D\big( \bar{\partial }\Hat{u}F_{\mu \nu }\hx^\nu \big)= 0~,
 \label{eq:eq_of_motionKK_x} \\
 &  \bar{\partial }D\Hat{u} = 0~,
 \label{eq:eq_of_motion_u} \\
 & \bar{\partial }D \Hat{v} + \half \bar{\partial }\big[\hx\cdot F \cdot D\hx \big] = 0~.
 \label{eq:eq_of_motion_v}
\end{align}
Let us solve these equations generally without use of any light-cone gauge, though original authors took the light-cone gauge. 
Since $\Hat{u}$ satisfies the free field equation (\ref{eq:eq_of_motion_u}), its solution has a form,
\begin{align}
 & \Hat{u}(s, \bar{s}) = \Hat{u}_R(s) + \Hat{u}_L(\bar{s})~.
 \label{eq:sol_sum}
\end{align}
By inserting Eq.(\ref{eq:sol_sum}) into Eq.(\ref{eq:eq_of_motionKK_x}), it reduces to
\begin{align}
 & \bar{\partial }D \hx  - \bar{\partial }\Hat{u}_L F\cdot D\hx= 0~.
 \label{eq:sol_u}
\end{align}
\indent Now, let us define a new variable
\begin{align}
 & \Hat{X}^\mu = \big(\exp{[-\Hat{u}_{L}(\bar{s})F]}\cdot \hx \big)^\mu ~, \quad \mu \neq 0,4~.
 \label{eq:new_X}
\end{align}
Then we have
\begin{align}
 & \bar{\partial }\Hat{X} = \exp{[-\Hat{u}_L(\bar{s})F]}\cdot [\bar{\partial }\hx -\bar{\partial }\Hat{u}_L F\cdot D\hx]
 \nonumber \\
 & D\Hat{X} = \exp{[-\Hat{u}_L(\bar{s})F]}\cdot D\hx~,
 \nonumber
\end{align}
so that
\begin{align}
 & \bar{\partial }D\Hat{X} = \exp{[-\Hat{u}_L(\bar{s})F]\cdot [\bar{\partial }D\hx -\bar{\partial }\Hat{u}_LF\cdot D\hx]}=0~.
 \label{eq:eq_of_X2}
\end{align}
because of Eq.(\ref{eq:sol_u}). This is nothing but a free equation $\Hat{X}$. So we can set as
\begin{align}
 & \hat{X}^\mu (s, \bar{s}) = \Hat{X}_R^\mu (s) + \Hat{X}_L^\mu (\bar{s})~, \quad \mu \neq 0,4~.
 \label{eq:solu_X_KK}
\end{align}
By using Eq.(\ref{eq:solu_X_KK}) the second term in Eq.(\ref{eq:eq_of_motion_v}) can be written as
\begin{align}
 & \bar{\partial }\big( \hx\cdot F \cdot D\hx \big) = \bar{\partial }\big(\Hat{X}\cdot F \cdot D\Hat{X} \big) 
    =\bar{\partial }\Hat{X}\cdot F \cdot D\Hat{X}
 \nonumber \\
 & =\bar{\partial }\Hat{X}_L\cdot F \cdot D\Hat{X}_R = \bar{\partial }D\big( \Hat{X}_L\cdot F \cdot \Hat{X}_R\big) ~.
 \label{eq:formura_X}
\end{align}
Hence, if we define
\begin{align}
 & \Hat{V} = \Hat{v} + \half \big( \Hat{X}_L\cdot F \cdot \Hat{X}_R\big)~,
 \label{eq:def_V}
\end{align}
then we get the free field equation
\begin{align}
 & \bar{\partial }D\Hat{V} = 0~.
 \label{eq:eq_of_motion_V}
\end{align}
Thus we have free field equations
\begin{align}
 & \bar{\partial }D\Hat{X}^\mu  = 0 \quad \mbox{for }\mu =u, V, 1, 2, 3~.
 \label{eq:eq_of_motion_V2}
\end{align}
Therefore, we can define a new Lagrangian
\begin{align}
 & \Hat{L}' = 2\bar{\partial }\Hat{X}_\mu D\Hat{X}^\mu \quad \mbox{for }\mu =u, V, 1, 2, 3, 5, 6, \cdots, d-1~.
 \label{eq:new_L}
\end{align}
from which Eqs.(\ref{eq:eq_of_motion_V2}) are derived. As for fields with components $\mu =3, 5, 6, \cdots, d-1$, they are completely free. 
We have exactly solved the KK model without use of any light-cone gauge. \\
\indent In complex variable notations, $\Hat{X}^{(\pm)}=( \Hat{X}^1 \pm i\Hat{X}^2)/\sqrt{2}$, 
Eq.(\ref{eq:new_X}) turns out to be of the form
\begin{align}
 & \Hat{X}^{(\pm)} = \exp{[\pm i\Hat{u}_L(\bar{s})B]}\hx^{(\pm)}~, \quad ( B=F^{12} )~.
 \label{eq:refrom_X}
\end{align}
In component fields of $\hx^{(\pm)}$, $x^{(\pm)}$ is periodic, $\psi ^\mu $ is anti-periodic (NS) or periodic (R) 
at boundaries $(\sigma =0, 2\pi )$, and $u_L(\Hat{s}+2\pi)=u_L(\Hat{s})+2\pi\alpha _0^u/\sqrt{2}$, 
so that we have boundary conditions 
for component fields of $\Hat{X}^{(\pm)}=X^{(\pm)}+i\theta \chi ^{(\pm)}/\sqrt{2} $ as
\begin{align}
 & X^{(\pm)}(\tau ,\sigma + 2\pi ) = \exp{\big( \pm 2\pi iB\alpha _0^u/\sqrt{2}\big)}X^{(\pm)}(\tau , \sigma ) 
                                   = \exp{\big( \pm 2\pi i\omega \big)}X^{(\pm)}(\tau , \sigma )~, 
 \label{eq:Xperiodicity2} \\
 & \chi ^{(\pm)}(\tau ,\sigma + 2\pi ) = \mp\exp{\big( \pm 2\pi i\omega \big)}\chi ^{(\pm)}(\tau , \sigma )~, 
 \label{eq:Psiperiodicity_chi2} 
\end{align}
where |and + signs in front of exponential functions stand for NS sector and Ramond sector, respectively, 
and $\omega $ is defined by $B\alpha _0^u/\sqrt{2}=N+\omega $, $N\in Z$, $0< \omega <1$ called the cyclotron frequency. \\
\indent The Lagrangian  $\hat{L}'$ can be rewritten as
\begin{align}
 & \Hat{L}' = 2\Big[ \Bar{\partial }\Hat{X}^{(+)} D\Hat{X}^{(-)} + \Bar{\partial }\Hat{X}^{(-)} D\Hat{X}^{(+)} \Big] 
+ 2 \sum_{\mu \neq 1,2} \partial \Hat{X}^{\mu}D\Hat{X}_{\mu}~.
 \label{eq:Lagragian8}
\end{align}
Integrating $\Hat{L}'$ over $\theta $, we have
\begin{align}
 & L' = 2\Big[ \Bar{\partial }X^{(+)} \partial X^{(-)} + \Bar{\partial }X^{(-)} \partial X^{(+)} 
  + i\half \big( \chi^{(+)}\Bar{\partial }\chi ^{(-)} + \chi^{(-)}\Bar{\partial }\chi ^{(+)} \big) \Big]  
  \nonumber \\
 & \quad \quad + 2 \sum_{\mu \neq 1,2} \big( \Bar{\partial }x^{\mu} \partial x_{\mu} + i\half \psi ^{\mu} \Bar{\partial }\psi _{\mu} \big)~,
 \label{eq:L_8}
\end{align}
where the sum runs over $\mu =u, V, 3, 5, 6, \cdots, d-1$. \\
\indent We conclude, therefore, that this KK heterotic string model is completely the same as that given before 
by Lagrangian (\ref{eq:intro_HL}). Hence, this model contains the same inconsistency as explained in Eq.(\ref{eq:physcond_L0_tild_L0}).
%
%
%
%
\section{Energy spectrum} \label{sec:9}
%
We consider the energy spectrum of (NS-NS) sector, as an example, in constant magnetic field. 
The right-moving Virasoro operator has the following form:
\begin{align}
 & L_{0}=\frac{\bar{p}^2}{2}+R_B+R_F, \quad \bar{p}^2\equiv \sum_{i=3}^{d-1} p_i^2 -p_0^2~,
  \label{eq:cond_L02} 
\end{align}
where
\begin{align}
 & R_B = R_B^{\mbox{free}} + \omega N_B +\omega N_0~,
 \label{eq:def_RB} \\
 & R_F = R_F^{\mbox{free}} + \omega N_F~,
 \label{eq:def_RF} \\
 & N_B = \sum_{n\geq 1} \big( N_n^{(-)} - N_n^{(+)}\big) ~, \quad N_n^{(\pm)} = \alpha _{-n}^{(\pm )}\alpha _{n}^{(\mp )}~,
 \nonumber \\
 & N_F = \sum_{r > 0} \big( N_r^{(-)} - N_r^{(+)}\big) ~, \quad N_r^{(\pm)} = b_{-r}^{(\pm )}b_{r}^{(\mp )}~,
 \label{eq:def_NF} \\
 & N_0^{(\pm)} = \alpha _{0}^{(-)}\alpha _{0}^{(+)}~.
 \nonumber 
\end{align}
Here $R$'s and $N$'s are all number operators. The left-moving Virasoro operator $\tilde {L}_{0}$ is composed of tilded ones, 
where
\begin{align}
 & \Tilde {N}_B = \sum_{n\geq 1} \big( \Tilde{N}_n^{(+)} - \Tilde{N}_n^{(-)}\big) ~, \quad \Tilde{N}_n^{(\pm)} 
      = \Tilde {\alpha }_{-n}^{(\pm )}\tilde {\alpha }_{n}^{(\mp )}~,
 \nonumber \\ 
 & \Tilde {N}_F = \sum_{r > 0} \big( \Tilde{N}_r^{(+)} - \Tilde{N}_r^{(-)}\big) ~, \quad \Tilde{N}_r^{(\pm)} 
      = \Tilde {\alpha }_{-r}^{(\pm )}\tilde {\alpha }_{r}^{(\mp )}~, 
 \label{eq:def_tilde_BF} \\
 & \Tilde {N}_0 = \Tilde {\alpha }_{0}^{(+)}\tilde {\alpha }_{0}^{(-)}~.
 \nonumber
\end{align}
\indent For physical states we have
\begin{align}
 & L_0 \approx \Tilde {L}_0 \approx \half(1-\omega )~.
 \label{eq:L0_phys}
\end{align}
Hence we get  $R_B+R_F\approx \Tilde {R}_B+\tilde {R}_F$, so that
\begin{align}
 & R_B^{\mbox{free}}+R_F^{\mbox{free}}\approx \Tilde {R}_B^{\mbox{free}}+\tilde {R}_F^{\mbox{free}}~,
 \label{eq:R_approx}
\end{align}

\begin{align}
 & N_B+N_F+N_0\approx \Tilde {N}_B+\tilde {N}_F+\tilde {N}_0~.
 \label{eq:N_approx}
\end{align}
\indent We also have $L_0+\tilde {L}_0\approx 1-\omega $ to yield
\begin{align}
 & \Hat{p}^2+R_B+R_F + \Tilde {R}_B+\Tilde {R}_F\approx 1-\omega ~.
 \label{eq:N_approx2}
\end{align}
This equation provides the Landau-like energy level for physical states. The ground state is the tachyon, 
$p_0^2-\sum_{i=3}^{d-1}p_i^2=-1+\omega <0$ , because of $0<\omega <1$. We do not discuss here the 
 GSO-projection, because this is irrelevant to our aim.
%
%
%
\section{Regularization criterion}\label{sec:10}
%
Some authors \cite{ref:Russo_AT2} have considered another kind of regularization based on the formula
 
\begin{align}
 & c_0= \lim_{e\to 0} \sum_{n=1}^\infty n^{-e}(n +  \omega ) = - \frac{1}{12} -\frac{\omega }{2}~.
 \label{eq:def_c_0}
\end{align}
This differs from Eq.\ref{eq:zeta_omega} in Appendix C by $-\omega ^2/2$, which is based on the generalized $\zeta $ function of Riemann defined as
\begin{align}
 & \zeta (s, a) = \sum_{n=0}^{\infty } \frac{1}{(n+a)^s}~, \quad 0 < a \leq 1~,
 \label{eq:zeta_func}
\end{align}
especially,
\begin{align}
 & \zeta (-1, \omega )-\omega  = \lim_{s \to -1} \sum_{n=1}^{\infty } (n+\omega )^{-s} = - \frac{1}{12} -\frac{\omega^2}{2}-\frac{\omega }{2}~.
 \label{eq:zeta_func-1}
\end{align}
\indent If the regularization (\ref{eq:def_c_0}) is used in the heterotic model, we have normal ordering constants 
\begin{align}
 & a = \half - \frac{\omega }{2}~, \quad \mbox{for the right-moving NS sector}~,
  \label{eq:regu_a} \\
 & \tilde {a} = 1 - \frac{\omega }{2}~, \quad \mbox{for the left-moving sector}~.
  \label{eq:regu_bar_a}
\end{align}
Since there is no $\omega ^2$ anomaly here, we have no inconsistency in the equation
\begin{align}
 & \big(L_0 - \tilde {L}_{0}-a + \tilde {a}\big)\ket{\mbox{phys.}}= \big(N - \tilde {N} +\half \big)\ket{\mbox{phys.}} =0~,
  \label{eq:phys_state_eq}
\end{align}
\indent As is shown in Appendix B, the regularization of Virasoro operators is performed by subtracting two infinite sums and making a shift 
in one of them. At first sight, therefore, this procedure seems to be ambiguous. In order to fix an appropriate regularization prescription, 
the above authors requires the Modular invariance for the partition function, which is assured by the first regularization prescription(\ref{eq:def_c_0})
\cite{ref:Russo_AT2}. \\
\indent However, we would like to stress that there is no ambiguity in the damping factor method. In Appendix B we have shown its uniqueness, 
i.e., the regularization never depends on any functional form of damping factors. \\
\indent In conclusion, the first regularization prescription is inconsistent with usual regularization prescriptions 
such as the operator product expansion in the text, contraction in Appendix A and the damping factor method in Appendix B. 
On the other hand, the second regularization prescription based on the generalized Zeta function of Riemann is consistent with these three regularizations. 
%
\section{Concluding remarks}\label{sec:11}
%
The heterotic string in a constant magnetic field can be solved exactly for the KK type and also for the minimal coupling type, 
without taking any light-cone gauge, as was shown in Secs.\ref{sec:7} and \ref{sec:8}. However, we pointed out that they 
include inconsistency coming from anomaly, which was explicitly explained in Eq.(\ref{eq:physcond_L0_tild_L0}). 
The bosonic string in the left-moving sector carries the anomaly,$(\omega -\omega ^2)/2$ , 
whereas the superstring (NS or R) in the right-moving sector carries 
the anomay,  $-\omega /2$ or 0. The $\omega ^2$ factor in Eq.(\ref{eq:physcond_L0_tild_L0}) can not be canceled by any eigenvalue 
of the number operator difference, $N-\tilde {N}$, which gives at most the first order of $\omega $. \\
\indent From this observation, we conclude that the exact solution in NSR superstring exists in each of combined sectors, 
(NS-NS), (NS-R) and (R-R), where there is no $\omega^2$ anomaly. Of course, (bosonic-bosonic) combination is allowed to have 
the exact solution in a constant magnetic field. \\
\indent We have also given the spectrum-generating algebra for our interacting system, which is necessary 
to construct actually physical states satisfying the Virasoro conditions. Finally it should be noted 
that our interacting models are equivalent to the completely free systems when the magnetic field times charge $qB$ 
takes integral values, $qB=N\in Z$.  
%
\begin{acknowledgments}
\indent We thank T. Okamura for useful discussions.@Thanks are also due to J. G. Russo and E. Kiritsis, 
who kindly informed us about their early works. Especially we owe a lot to J. G. Russo and A. A. Tseytlin 
for continuous long time discussions and valuable comments.
\end{acknowledgments}
\appendix
\section{Calculation of anomalies based on contraction}\label{sec:appendA}
%
The most simple method to obtain anomalies for relevant parts is to calculate contractions of $\big[\, L_m\,,~L_n\, \big]$, or 
$\big\{\, G_r\,,~G_s\, \big\}$.
 For the bosonic case we have
\begin{align}
 & \langle~\big[\, L_m\,,~L_n\, \big]~\rangle = \sum_{k,l} \langle~:\alpha _k^{(+)}\alpha _{m-k}^{(-)}::\alpha _l^{(+)}\alpha _{n-l}^{(-)}: 
- :\alpha _l^{(+)}\alpha _{n-l}^{(-)}::\alpha _k^{(+)}\alpha _{m-k}^{(-)}:~\rangle
 \nonumber \\
 & \quad \quad = \sum_{k,l} \langle~\alpha _k^{(+)}\alpha _{n-l}^{(-)}~\rangle \langle~\alpha _{m-k}^{(-)}\alpha _{l}^{(+)}~\rangle 
 - \sum_{k,l} \langle~\alpha _l^{(+)}\alpha _{m-k}^{(-)}~\rangle \langle~\alpha _{n-l}^{(-)}\alpha _{k}^{(+)}~\rangle
 \label{eq:VEV_LL}
\end{align}
where contractions are defined as
\begin{align}
 & \langle~\alpha _m^{(+)}\alpha _n^{(-)}~\rangle = \delta _{m+n,0} \theta _{m\geq 0}(m+\omega )~,
  \nonumber \\
 & \langle~\alpha _m^{(-)}\alpha _n^{(+)}~\rangle = \delta _{m+n,0} \theta _{m> 0}(m-\omega )~,
 \label{eq:VEV_alpha_alpha} \\
 & \quad 0 < \omega < 1~, \quad  
   \theta _\Gamma = \left\{
    \begin{array}{rl}
     1,& \quad \mbox{if $\Gamma$ is true} \\
     0,& \quad \mbox{if $\Gamma$ is false}
    \end{array}\right.
 \nonumber 
\end{align}  
These equations give a finite sum so that we have the unique anomaly $A_m^B$ with $\omega ^2$ term.
\begin{align}
 & \langle~\big[\, L_m\,,~L_n\, \big]~\rangle = \delta _{m+n,0} A_m^B~,
 \nonumber \\
 & \quad A_m^B = \sum_{k=0}^{m-1} ( m-k-\omega )(k+\omega ) 
  = \frac{1}{6}m(m^2-1) +m\omega (1-\omega )~.
 \label{eq:VEV_LL2}
\end{align}
\indent In the same way, for the superstring case we have
\begin{align}
 & \langle~G_r~G_s~\rangle = \sum_{m,n} \langle~b_{r-m}^{(+)}\alpha _{m}^{(-)}b_{s-n}^{(-)}\alpha _{n}^{(+)} 
  + b_{r-m}^{(-)}\alpha _{m}^{(+)}b_{s-n}^{(+)}\alpha _{n}^{(-)}~\rangle
 \nonumber \\
 & \quad \quad \quad = \sum_{m,n} \langle~b_{r-m}^{(+)}\alpha _{s-n}^{(-)}~\rangle \langle~\alpha _{m}^{(-)}\alpha _{n}^{(+)}~\rangle 
 + \sum_{m,n} \langle~b_{r-m}^{(-)}b_{s-n}^{(+)}~\rangle \langle~\alpha _{m}^{(+)}\alpha _{n}^{(-)}~\rangle ~.
 \label{eq:VEV_GG}
\end{align}
Here contractions for fermionic operators have been defined as
where contractions are defined as
\begin{align}
 & \langle~b_{r-m}^{(+)}b_{s-n}^{(-)}~\rangle = \langle~b_{r-m}^{(-)}b_{s-n}^{(+)}~\rangle = \delta _{t+s,m+n} \theta _{r-m>0}~,
  \label{eq:VEVbb} 
\end{align}  
so that we get a finite sum for each contraction to yield
\begin{align}
 & \langle~G_r~G_s~\rangle = \delta _{r+s,0}B_r~,
 \label{eq:VEV_GG2} \\
 & \quad B_r = r^2 - \frac{1}{4} + \omega ~.
 \nonumber
\end{align}
The corresponding anomaly of $\big[\, L_m\,,~L_n\, \big]$ will be obtained by using formulae,
\begin{align}
 & \big[\, L_m\,,~L_{r+s}\, \big] = \half \big[\, L_m\,,~\big\{\, G_r\,,~G_s\, \big\}\, \big] ~,
 \label{eq:commutation_LL}
\end{align}
and
\begin{align}
 & \big[\, L_m\,,~G_{r}\, \big] = \big(\half m - r\big) G_{m+r}~. 
 \label{eq:commutation_LG}
\end{align}
The result is
\begin{align}
 & A_m^s = \frac{1}{4}m\big(m^2 - 1\big)+m\omega ~, 
 \label{eq:A_m_s}
\end{align}
without $\omega ^2$ term.
%
\section{Uniqueness of anomalies based on the damping factor method}\label{sec:appendB}
Commutation relations for bosonic string are given by
\begin{align}
 & \big[\, \alpha _m^{(+)}\,,~\alpha _n^{(-)}\, \big] = (m+\omega )\delta _{m+n,0} g_{m}~, 
 \label{eq:commutation_alpha_alpha_B} \\
 & \big[\, \alpha _m^{(-)}\,,~\alpha _n^{(+)}\, \big] = (m-\omega )\delta _{m+n,0} g_{-m}~.
 \nonumber
\end{align}
where a damping factor $g_m$ is inserted. Its detailed functional form is irrelevant, 
but it should be set as $g_m=1$ when series is a finite sum. \\
\indent The relevant two dimensional Virasoro operator is given by
\begin{align}
 &  L_n = \half \sum_{k}:\big( \alpha _{k}^{(+)} \alpha _{n-k}^{(-)} + \alpha _{k}^{(-)} \alpha _{n-k}^{(+)} \big) : 
   =  \sum_{k} :\alpha _{k}^{(+)} \alpha _{n-k}^{(-)}: ~.
  \label{eq:L_nB}
\end{align}
Then we have
\begin{align}
 & \big[\, L_m\,,~L_{-n}\, \big] = \sum_{k,l}\big[ \alpha _{k}^{(+)} \alpha _{m-k}^{(-)} \,,~
		\alpha _{l}^{(+)} \alpha _{n-l}^{(-)} \, \big]
   \nonumber \\
 & \quad = \sum_{k,l}\Big\{\big[ \alpha _{m-k}^{(-)} \,,~\alpha _{l}^{(+)}\, \big]\alpha _{k}^{(+)} \alpha _{n-l}^{(-)} 
        + \big[ \alpha _{k}^{(+)} \,,~\alpha _{n-l}^{(-)} \, \big]\alpha _{l}^{(+)} \alpha _{m-k}^{(-)} \Big\} 
   \label{eq:commutaion_LL_B} \\
 & \quad = \sum_{k,l} \big\{ \delta _{m-k+l,0}(m-k-\omega )g_{-m+k}\alpha _k^{(+)} \alpha _{n-l}^{(-)}
 + \delta _{k+n-l,0}(k+\omega )g_{k}\alpha _l^{(+)} \alpha _{m-k}^{(-)} \big\} 
 \nonumber \\
 & \quad = \sum_k \big\{ (m-k-\omega )g_{-m+k}\alpha _k^{(+)} \alpha _{m+n-k}^{(-)} + (k+\omega )g_{k}\alpha _{n+k}^{(+)} \alpha _{m-k}^{(-)} \big\} 
 \nonumber \\
 & \quad = (m-n)L_{m+n} + \delta _{m+n,0}A_m^B~,
 \nonumber \\
 & \quad \quad A_m^B = \sum _{k\geq m} (m-k-\omega )(k+\omega )g_{-m+k} g_k+\sum _{k\geq m} (k+\omega )(k-m+\omega )g_k g_{k-m}~.
 \label{eq:def_A_m_B}
\end{align}
When $m>0$, it follows that
\begin{align}
 & A_m^B = \sum _{0\leq  k < m} (m-k-\omega )(k+\omega )g_k g_{k-m}
 \nonumber \\
 & \quad = \sum _{k=0}^{m-1} (m-k-\omega )(k+\omega )
 \label{eq:A_m^B_Value} \\
 & \quad = \frac{1}{6}m(m^2-1) + m\omega (1-\omega )~.
 \nonumber
\end{align}
Here the damping factors, $g_k, g_{k-m}$, have been set as unity, since the series is finite. The same is true when $m\leq 0$. 
The result agrees with Eq.(\ref{eq:VEV_LL2}) with $\omega ^2$ term. \\
\indent For the NS string case, commutation relations of fermionic mode operators are given by
\begin{align}
 & \big\{\, b_r^{(+)}\,,~b_s^{(-)}\, \big\} = \delta _{r+s,0} \gamma _{r}~, 
 \quad 
 \big\{\, b_r^{(-)}\,,~b_s^{(+)}\, \big\} = \delta _{r+s,0} \gamma _{-r}~.
 \label{eq:commutation_b_b_B}
\end{align}
where $\gamma _{r}$ is the damping factor. By using Eqs.(\ref{eq:commutation_alpha_alpha_B}) and (\ref{eq:commutation_b_b_B}), let us calculate the anti-commutator
\begin{align}
 & \big\{\, G_r\,,~G_{s}\, \big\} = \sum_{m,n}\big\{ b_{r-m}^{(+)} \alpha _{m}^{(-)}+b_{r-m}^{(-)} \alpha _{m}^{(+)} \,,~
		b_{s-n}^{(+)} \alpha _{n}^{(-)}+b_{s-n}^{(-)} \alpha _{n}^{(+)} \, \big\}
   \nonumber \\
 & \quad = \sum_{m,n}\Big[\big\{ b_{r-m}^{(+)} \alpha _{m}^{(-)} \,,~b_{s-n}^{(-)} \alpha _{n}^{(+)}\, \big\}
        + \big\{ b_{r-m}^{(-)} \alpha _{m}^{(+)} \,,~b_{s-n}^{(+)} \alpha _{n}^{(-)} \, \big\}\Big] 
   \nonumber \\
 & \quad = \sum_{m,n}\Big[\big\{ b_{r-m}^{(+)}  \,,~b_{s-n}^{(-)} \, \big\}\alpha _{m}^{(-)}\alpha _{n}^{(+)}
		+\big[ \alpha _{n}^{(+)} \,,~ \alpha _{m}^{(-)}\, \big] b_{s-n}^{(-)}b_{r-m}^{(+)} 
 \nonumber \\
 & \quad \quad + \big\{ b_{r-m}^{(-)}  \,,~b_{s-n}^{(+)} \, \big\}\alpha _{m}^{(+)}\alpha _{n}^{(-)} 
          +\big[ \alpha _{n}^{(-)} \,,~ \alpha _{m}^{(+)} \, \big]b_{s-n}^{(+)}b_{r-m}^{(-)} \Big] 
  \label{eq:commutaion_G_G_B} \\
 & \quad = \sum_{m,n}\big[ \delta _{r+s-m-n,0}\gamma _{r-m}\alpha _m^{(-)} \alpha _{n}^{(+)}
		+ \delta _{m+n,0}(n+\omega )g_n b_{s-n}^{(-)}b_{r-m}^{(+)}
    \nonumber \\
 & \quad \quad + \delta _{r+s-m-n,0}\gamma _{-r+m}\alpha _m^{(+)} \alpha _{n}^{(-)} 
        + \delta _{m+n,0} (n-\omega )g_{-n} b_{s-n}^{(+)}b_{r-m}^{(-)} \big\} 
  \nonumber \\
 & \quad = \sum_{m}\big[ \gamma _{r-m}\alpha _m^{(-)} \alpha _{r+s-m}^{(+)}
		+ (-m+\omega )g_{-m} b_{s+m}^{(-)}b_{r-m}^{(+)}
    \nonumber \\
 & \quad \quad + \gamma _{-r+m}\alpha _m^{(+)} \alpha _{r+s-m}^{(-)} 
        + (-m-\omega )g_{m} b_{s+m}^{(+)}b_{r-m}^{(-)} \big]~. 
  \nonumber
\end{align}
Accordingly we get
\begin{align}
 & \big\{\, G_r\,,~G_{s}\, \big\} = 2L_{r+s} + \delta _{r+s,0}B_r~,
   \label{eq:commution_G_GB2} \\
 & \quad B_r = \sum_{m>0}\gamma _{r-m}g_{-m}(m-\omega ) - \sum_{m>r}\gamma _{r-m}g_{-m}(m-\omega )
   \nonumber \\
 & \quad \quad + \sum_{m\geq 0}\gamma _{-r+m}g_{m}(m+\omega ) - \sum_{m>r}\gamma _{-r+m}g_{m}(m+\omega )~.
   \nonumber
\end{align}
When $r>0$, it follows that
\begin{align}
 & B_r = \sum_{m=1}^{r-1/2} \gamma _{r-m}g_{-m}(m-\omega ) - \sum_{m=0}^{r-1/2} \gamma _{-r+m}g_{m}(m+\omega )~.
   \label{eq:Cal_B_r}
\end{align}
Since this is a finite sum, one can set as $\gamma _r= g_m=1$ to yield 
\begin{align}
 & B_r = 2\sum_{m=1}^{r-1/2} m+\omega=r^2 - \frac{1}{4} +\omega ~.
   \label{eq:Cal_B_r2}
\end{align}
The same is true even when $r<0$. \\
\indent In this calculation we do not need any symmetry of $\gamma _r$, $g_m$.with respect to $\pm r$ and $\pm m$. \\
\indent In conclusion this regularization never depends on any functional form of the damping factor. 
Therefore, there is no ambiguity in this method, giving a unique result.
%
\section{Regularization by means of the generalized Zeta function of Riemann}\label{sec:appendC}
The generalized $\zeta $ function of Riemann is defined as
\begin{align}
 & \zeta (s, a) = \sum_{n=0}^{\infty } \frac{1}{(n+a)^{s}}~, \quad 0 < a \leq 1~.
 \label{eq:zeta_C}
\end{align}
Especially we have
\begin{align}
 & \zeta (-1, \omega )-\omega  = \sum_{n=1}^{\infty } (n+\omega )=-\frac{1}{12}-\frac{\omega ^2}{2} - \frac{\omega }{2}~, 
 \label{eq:zeta_omega}
\end{align}
and
\begin{align}
 & \sum_{n=1}^{\infty } (n-\omega )=-\frac{1}{12}-\frac{\omega ^2}{2} + \frac{\omega }{2}~, 
 \label{eq:zeta_omega2}
\end{align}
in the region $0 < \omega  < 1$. \\ 
\indent By using these formulae let us calculate normal ordering constants for NS sector. The un-normal ordered Virasoro 0-th operator is given by
\begin{align}
 &  L_0 = \half \sum_{n}\big( \alpha _{-n}^{(+)} \alpha _{n}^{(-)} + \alpha _{-n}^{(-)} \alpha _{n}^{(+)} \big) 
        +\half \sum_{n} \sum_{\mu ,\nu \neq 1,2} \eta _{\mu \nu } \alpha _{-n}^{\mu } \alpha _{n}^{\nu } 
   \label{eq:def_L_n_C} \\
 &  \quad \quad  +\half \sum_{r} \big( r-\omega \big)  b_{-r}^{(+)}b_{r}^{(-)} 
        +\half \sum_{r}\big( r+\omega \big) b_{-r}^{(-)}b_{r}^{(+)} 
   +\half \sum_{r}\sum_{\mu ,\nu \neq 1,2} \eta _{\mu \nu } r b_{-r}^{\mu } b_{r}^{\nu }~,
   \nonumber
\end{align}
with
\begin{align}
 & \big[\, {\alpha }_n^{(\pm)}\,,~{\alpha }_{-n}^{(\mp)}\, \big] = (n\mp \omega )~,
\label{eq:2}  \\
 & \big\{\, b_r^{(+)}\,,~b_{s}^{(-)}\, \big\} = \delta _{r+s,0}~.
\end{align}
%
\subsection{Bosonic sector}
%
The bosonic sector becomes
\begin{align}
 &  L_0^B = \half \sum_{n\geq 1}\big( \alpha _{-n}^{(+)} \alpha _{n}^{(-)} + \alpha _{-n}^{(-)} \alpha _{n}^{(+)} \big) 
		+ \half \sum_{n\geq 1}\big( \alpha _{n}^{(+)} \alpha _{-n}^{(-)} + \alpha _{n}^{(-)} \alpha _{-n}^{(+)} \big) 
        +\half \sum_{n\geq 1} \sum_{\mu ,\nu \neq 1,2} \eta _{\mu \nu }\big( \alpha _{-n}^{\mu } \alpha _{n}^{\nu } + \alpha _{n}^{\mu } \alpha _{n}^{\nu } \big)
   \nonumber \\
 &  \quad \quad  + \half \big( \alpha _{0}^{(+)} \alpha _{0}^{(-)} + \alpha _{0}^{(-)} \alpha _{0}^{(+)} \big) 
	    + \half \sum_{\mu ,\nu \neq 1,2} \eta _{\mu \nu } p^{\mu } p^{\nu }
   \nonumber \\
 & \quad = \sum_{n\geq 0}\big( \alpha _{-n}^{(+)} \alpha _{n}^{(-)} + \alpha _{-n}^{(-)} \alpha _{n}^{(+)} \big)
		+ \half \sum_{n\geq 1} (n+\omega ) + \half \sum_{n\geq 1} (n-\omega ) 
	    + \sum_{n\geq 1} \sum_{\mu ,\nu \neq 1,2} \eta _{\mu \nu } \alpha _{-n}^{\mu } \alpha _{n}^{\nu }
  \nonumber \\
 & \quad \quad   + \half\sum_{n\geq 1} \sum_{\mu ,\nu \neq 1,2} \eta _{\mu \nu } \eta ^{\mu \nu } n 
 		+ \alpha _{0}^{(-)} \alpha _{0}^{(+)} +\half \omega + \half {p'}^2
\end{align}
From Eqs.(\ref{eq:zeta_omega}) and (\ref{eq:zeta_omega2}) we have
\begin{align}
 & \half \sum_{n\geq 1} (n+\omega ) + \half \sum_{n\geq 1} (n-\omega ) = - \frac{1}{12} - \half \omega ^2~.
  \nonumber
\end{align}
Also
\begin{align}
 & \half\sum_{n\geq 1} \sum_{\mu ,\nu \neq 1,2} \eta _{\mu \nu } \eta ^{\mu \nu } n = - \frac{1}{4} \quad 
    \mbox{(only for transverse sectors)}~.
 \nonumber
\end{align}
Then, totally we get
\begin{align}
 & -a_B = -\frac{1}{3} - \half \omega ^2 + \half \omega ~.
 \label{eq:a_B_C}
\end{align}
It may be remarkable that the $\omega ^2$ anomaly appears in the bosonic sector.
%
\subsection{Fermionic sector}
%
The fermionic sector becomes
\begin{align}
 &  L_0^F = \half \sum_{r\geq 1/2} \big( r-\omega \big)  b_{-r}^{(+)}b_{r}^{(-)}+\half \sum_{r\geq 1/2} \big( -r-\omega \big)  b_{r}^{(+)}b_{-r}^{(-)} 
  \nonumber \\
 & \quad \quad    +\half \sum_{r\geq 1/2}\big( r+\omega \big) b_{-r}^{(-)}b_{r}^{(+)} + +\half \sum_{r\geq 1/2}\Big(-r+\omega \big) b_{r}^{(-)}b_{-r}^{(+)} 
    \nonumber \\
 & \quad \quad  +\half \sum_{r\geq 1/2}\sum_{\mu ,\nu \neq 1,2}\eta _{\mu \nu } r \big( b_{-r}^{\mu } b_{r}^{\nu } - b_{r}^{\mu } b_{-r}^{\nu } \big)
   \label{eq:L_0_F_8} \\
 & \quad = \sum_{r\geq 1/2} \big( r-\omega \big)  b_{-r}^{(+)}b_{r}^{(-)} + \half \sum_{r\geq 1/2} \big( -r-\omega \big)
   \nonumber \\
 & \quad \quad + \sum_{r\geq 1/2}\big( r+\omega \big) b_{-r}^{(-)}b_{r}^{(+)} + \half \sum_{r\geq 1/2} \big( -r+\omega \big)
   \nonumber \\
 & \quad \quad + \sum_{r\geq 1/2}\sum_{\mu ,\nu \neq 1,2}\eta _{\mu \nu } r b_{-r}^{\mu } b_{r}^{\nu }
    - \half \sum_{r\geq 1/2}\sum_{\mu ,\nu \neq 1,2}\eta _{\mu \nu } \eta ^{\mu \nu } r~. 
   \nonumber
\end{align}
By using the formulae
\begin{align}
 & \half \sum_{r\geq 1/2} \big( -r-\omega \big) = -\frac{1}{48} + \frac{\omega ^2}{4}~,
 \nonumber \\
 & \half \sum_{r\geq 1/2} \big( -r+\omega \big) = -\frac{1}{48} + \frac{\omega ^2}{4}~,
 \label{eq:sum_C} \\
 & - \half \sum_{r\geq 1/2}\sum_{\mu ,\nu \neq 1,2}\eta _{\mu \nu } \eta ^{\mu \nu } r = - \frac{1}{8}~, \quad \mbox{only for transverse sectors}
 \nonumber
\end{align}
we get
\begin{align}
 & - a_F = -\frac{1}{6} + \frac{\omega ^2}{2}~.
 \label{eq:a_F_C}
\end{align}
The total sum is given by
\begin{align}
 & a = a_B + a_F = \frac{1-\omega }{2}~.
 \label{eq:a_F_C2}
\end{align}
This agrees with Eq.(\ref{eq:d-}). It is remarkable that the $\omega ^2$ anomaly is cancelled out in $a$. \\
\indent For the heterotic sring model we have
\begin{align}
 & a = \frac{1-\omega }{2}~, \quad \mbox{for the right-moving NS sector}~,
 \label{eq:a_right-moving_C} \\
 & \tilde {a} = 1 - \frac{\omega -\omega^2 }{2}~, \quad \mbox{for the left-moving sector}~.
 \label{eq:a_left-moving_C}   
\end{align}
%
%

\end{document}